\documentclass[a4paper,11pt]{article}
\usepackage{pos}
\usepackage{graphicx}
\usepackage{subfigure}
\usepackage{wrapfig}
\usepackage{caption}
\usepackage{mwe}
\usepackage{amsmath}
\usepackage{hyperref}

\title{Simulation of $e^+e^-$ annihilation with quark spin effects}

\author*[a]{A. Kerbizi}
\author[b]{L. L\"onnblad}
\author[a]{A. Martin}

\affiliation[a]{Dipartimento di Fisica, Universitá degli Studi di Trieste and INFN Sezione di Trieste,\\
Via Valerio 2, 34127 Trieste, Italy}

\affiliation[b]{Department of Physics,\\
Box 118, 221 00 Lund, Sweden}

\emailAdd{albi.kerbizi@ts.infn.it}

\def\GeV{\rm GeV}
\def\q{q}
\def\qp{\q'}
\def\qbar{\bar{q}}
\def\qbarp{\bar{q}'}
\def\fL{f_{\rm L}}
\def\thetaLT{\theta_{\rm LT}}
\def\PT{P_{\rm T}}
\def\PTi{P_{i\rm T}}
\def\PTa{P_{1\rm T}}
\def\PTb{P_{2\rm T}}
\def\phiH{\phi_{12}}
\def\n{\hat{\textbf{n}}}
\def\T{\hat{\textbf{T}}}
\def\Q{\hat{\textbf{Q}}}
\def\X{X}
\def\Xa{\X_1}
\def\Xb{\X_2}
\def\Xi{\X_i}

\def\AUL{A_{12}^{\rm UL}}
\def\pmin{\textbf{p}_-}
\def\pp{\textbf{p}_+}

\def\kqp{k'}
\def\kVec{\textbf{k}}
\def\kbarVec{\bar{\textbf{k}}}
\def\xq{\hat{\textbf{x}}_q}
\def\yq{\hat{\textbf{y}}_q}
\def\zq{\hat{\textbf{z}}_q}
\def\xqbar{\hat{\textbf{x}}_{\qbar}}
\def\yqbar{\hat{\textbf{y}}_{\qbar}}
\def\zqbar{\hat{\textbf{z}}_{\qbar}}

\def\Iq{1^q}

\def\sigmaXq{\sigma_x^q}
\def\sigmaYq{\sigma_y^q}
\def\sigmaZq{\sigma_z^q}
\def\Iqbar{1^{\qbar}}

\def\sigmaXqbar{\sigma_x^{\qbar}}
\def\sigmaYqbar{\sigma_y^{\qbar}}
\def\sigmaZqbar{\sigma_z^{\qbar}}

\def\Trqq{\rm{Tr}_{q\qbar}}

\def\kt{\textbf{k}_{\rm T}}
\def\kpt{\textbf{k}'_{\rm T}}
\def\pt{\textbf{p}_{\rm T}}

\def\ktbar{\bar{\textbf{k}}_{\rm T}}
\def\kptbar{\bar{\textbf{k}}'_{\rm T}}
\def\Pt{\textbf{P}_{\rm T}}

\def\V{\textbf{V}}
\def\T{\textbf{T}}
\def\Tr{\rm Tr}

\abstract{
The quark spin effects are introduced for the first time in the string fragmentation routine of the Pythia 8 Monte Carlo event generator for the simulation of $e^+e^-$ annihilation to hadrons. To describe the spin effects the string+${}^3P_0$ model of polarized hadronization with emissions of pseudoscalar and vector mesons is used. The spin effects are activated in the generator by extending the StringSpinner package, previously applied to the simulation of deep inelastic scattering off a polarized nucleon.
The generator is used to carry out simulations of $e^+e^-$ annihilation at the center of mass energy of $10.6\,\GeV$. The Collins asymmetry for back-to-back pion pairs is evaluated and compared to the asymmetry as measured by the BELLE experiment. A satisfactory agreement is found.}

\FullConference{25th International Spin Physics Symposium (SPIN 2023)\\
 24-29 September 2023\\
 Durham, NC, USA\\}

\begin{document}
\maketitle

\section{Introduction}
According to the factorization theorem \cite{Collins:1981uk}, the annihilation reaction $e^+e^-\rightarrow \rm hadrons$ can be factorized in an elementary hard interaction $e^+e^-\rightarrow \q\qbar$ where a quark pair $\q\qbar$ is produced, and the subsequent hadronization of $\q$ and $\qbar$ in the final state hadrons. The latter is a soft QCD process described usually by fragmentation functions (FFs), which encode the dynamics behind the conversion of quarks and gluons in hadrons. Among these, the class of spin dependent FFs is particularly releveant as they allow to access the partonic transverse spin structure of the nucleons by a combined phenomenological analysis of data from semi-inclusive deep inelastic scattering (SIDIS) and $e^+e^-$ annihilation (for a review see, e.g., Ref. \cite{Anselmino:2020vlp}). An example is the extraction of the transverse spin distribution of quarks in a transversely polarized nucleon described by the transversity parton distribution function, which requires the knowledge of the Collins FF. The latter describes the fragmentation of a transversely polarized quark in an unpolarized hadron, and information on the FF is obtained from the Collins asymmetries for back-to-back hadrons in $e^+e^-$ annihilation. The asymmetry was measured by the BELLE \cite{Belle:2011cur,Belle:2019nve}, BABAR \cite{BaBar:2013jdt,BaBar:2015mcn} and BESIII \cite{BESIII:2015fyw} experiments. The transversity PDF and the Collins FF were extracted by phenomenological fits of the Collins asymmetries in SIDIS and in the $e^+e^-$ annihilation by different groups.

An alternative approach to the phenomenological extractions of the Collins FF is the modeling of hadronization and its implementation in Monte Carlo event generators. In this work we implement the quark spin effects in the Pythia 8.3 event generator \cite{Bierlich:2022pfr} for the simulation of $e^+e^-$ annihilation to hadrons. To describe the quark spin effects in hadronization we use the string+${}^3P_0$ model of Ref. \cite{Kerbizi:2021gos}, which was recently extended to the simulation of the string fragmentation process of a $\q\qbar$ pair with entangled spin states \cite{Kerbizi:2023luv}. The implementation of the model in the Pythia generator is achieved by developing further the StringSpinner package \cite{Kerbizi:2021pzn,Kerbizi:2023cde}, which presently implements the string+${}^3P_0$ model in Pythia hadronization for the simulation of polarized DIS. The implementation of the string+${}^3P_0$ model in Pythia for $e^+e^-$ annihilation is described in Sec. \ref{sec:implementation}. The new StringSpinner package is used to simulate $e^+e^-$ events at the center of mass energy (c.m.s) $\sqrt{s}=10.6\,\GeV$, and the resulting Collins asymmetries for back-to-back hadrons is evaluated. The simulation results as well as the comparison with data are shown in Sec. \ref{sec:results}. The conclusions are given in Sec. \ref{sec:conclusions}.


\section{Implementation of the string+${}^3P_0$ model in Pythia for $e^+e^-$}\label{sec:implementation}
\begin{figure}[tbh]
\centering
\begin{minipage}[b]{0.45\textwidth}
\hspace{-2.0em}
\includegraphics[width=1.0\textwidth]{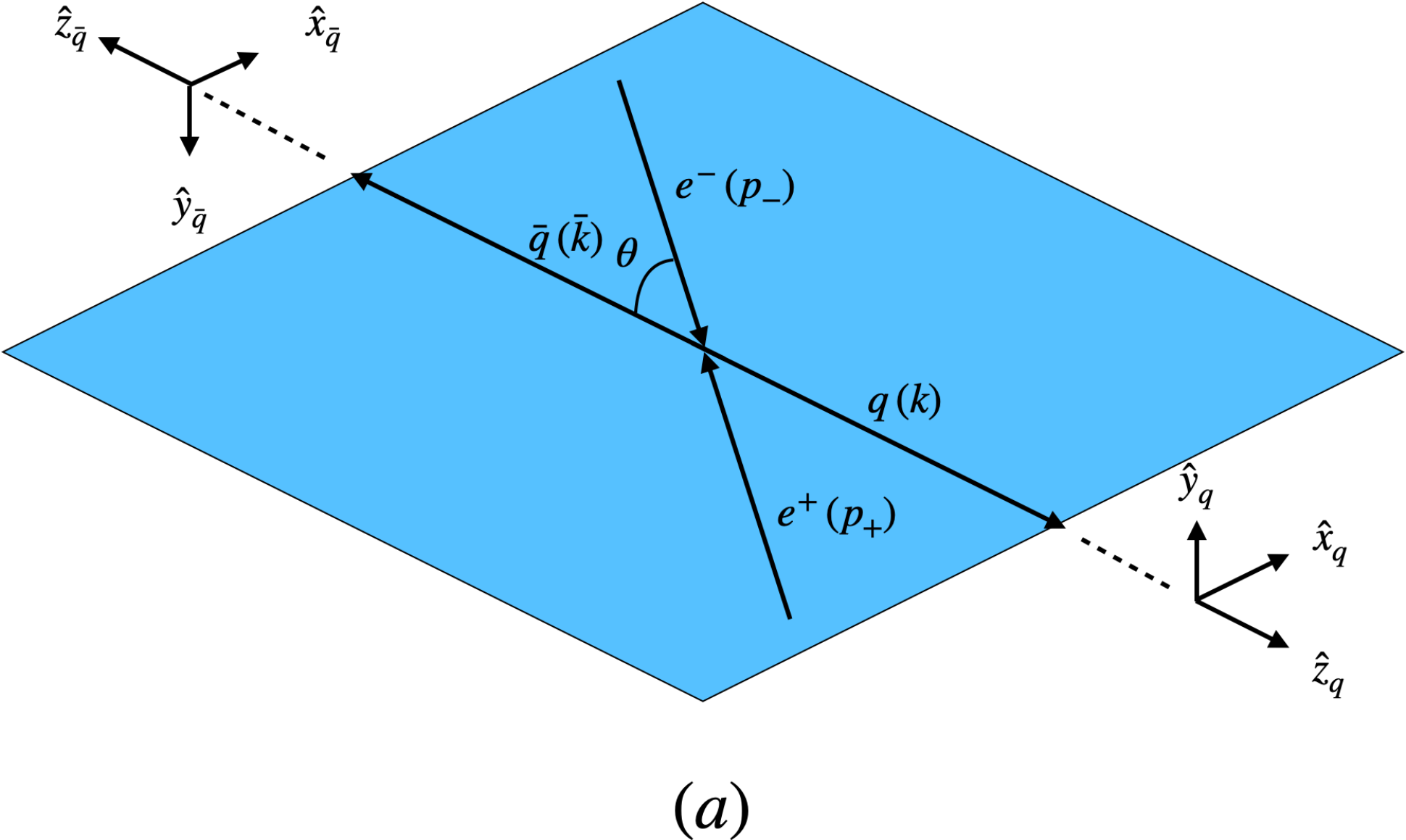}
\end{minipage}
\begin{minipage}[b]{0.45\textwidth}
\hspace{1.0em}
\includegraphics[width=1.0\textwidth]{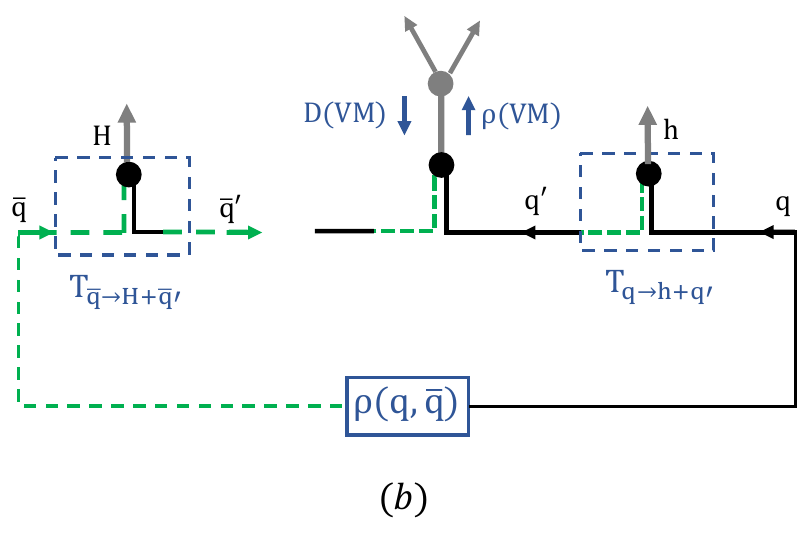}
\end{minipage}
\caption{Kinematics of the $e^+e^-\rightarrow q\qbar$ in the c.m.s (a). Representation of the polarized string fragmentation process in StringSpinner (b).}
\label{fig:kinematics}
\end{figure}

To begin the simulation, we let Pythia generate the kinematics of the hard reaction $e^+e^-\rightarrow \q\qbar$. The process is considered at leading order, and parton showers have been switched off. The kinematics is shown in Fig. \ref{fig:kinematics}a, where $\theta$ is angle between the momentum $\pmin$ of $e^-$ and the momentum $\kVec$ of $\q$. The momenta of $e^+$ and $\qbar$ are indicated by $\pp$ and $\kbarVec$, respectively. Following Ref. \cite{Kerbizi:2023luv} \footnote{The quark $q$ ($q'$) and the antiquark $\bar{q}$ ($\bar{q}'$) in this work are indicated by $q_1$ ($q_2$) and $\bar{q}_1$ ($\bar{q}_2$), respectively, in Ref. \cite{Kerbizi:2023luv}.}, we define the quark helicity frame (QHF) and antiquark helicity frame (AHF) by the set of axes $\lbrace \xq, \yq,\zq\rbrace$ and $\lbrace \xqbar, \yqbar, \zqbar\rbrace$, respectively. The axes of the QHF are obtained by $\zq=\kVec/|\kVec|$, $\yq=\pmin\times\zq/|\pmin\times \zq |$ and $\xq= \yq\times \zq$. The axes of the AHF are obtained analogously by using $\kbarVec$ instead of $\kVec$. The QHF and AHF are also shown in Fig. \ref{fig:kinematics}a.

Before starting the fragmentation of the string stretched between $\q$ and $\qbar$, the joint spin density matrix $\rho(\q,\qbar)$ is set up. It implements the correlations between the (entangled) spin states of $\q$ and $\qbar$. Neglecting quark masses, the joint spin density matrix reads \cite{Kerbizi:2023luv}
\begin{equation}\label{eq:rho}
\rho(\q,\qbar)=\left[\Iq\otimes \Iqbar - \sigmaZq\otimes \sigmaZqbar + \frac{\sin^2\theta}{1+\cos^2\theta}\,(\sigmaXq\otimes \sigmaXqbar + \sigmaYq\otimes \sigmaYqbar)\right]/4,
\end{equation}
where $\sigma_{\nu}^{\q(\qbar)}$ indicates the Pauli matrix along the axis $\nu=0,x,y,z$ ($0$ referring to the identity matrix) in the QHF (AHF).

Pythia then starts the fragmentation of the $\q-\qbar$ string by selecting randomly emissions of hadrons from the $\q$ and $\qbar$ sides of the string. As can be seen in Fig. \ref{fig:kinematics}b, emissions from the $\q$ side are viewd as splittings $\q\rightarrow h + \qp$, where $h$ is the emitted hadron with four momentum $p$ and $\qp$ is the leftover quark with four-momentum $\kqp$. The transverse momenta of $\q$, $h$ and $\qp$ with respect to the string axis are defined to be $\kt$, $\pt$ and $\kpt$, respectively. Momentum conservation implies $\kpt=\kt-\pt$.
Analogously, emissions from the $\qbar$ side are viewed as splittings $\qbar\rightarrow H + \qbarp$, with $H$ being the emitted hadron and $\qbarp$ the leftover antiquark. The transverse momenta of $\qbar$, $H$ and $\qbarp$ with respect to the string axis are defined as $\ktbar$, $\Pt$ and $\kptbar$, respectively. They are related by $\kptbar = \ktbar-\Pt$. 

Following the previous implementation of StringSpinner \cite{Kerbizi:2023cde} only the productions of pseudoscalar (PS) mesons and vector mesons (VMs) are activated. Then, if the splitting is taken from the $\q$ side, the hadron $h$ is accepted with a probability
\begin{equation}\label{eq:wh}
w_h(\pt;\kt)=\Trqq\left[\T_{\qp,h\,\q}\,\rho(\q,\qbar)\T^{\dagger}_{\qp,h\,\q}\right]/\Trqq\left[\T_{\qp,h\,\q}\,\T^{\dagger}_{\qp,h\,\q}\right],
\end{equation}
where $\T_{\qp,h,\q}=T_{\qp,h,\q}\otimes \Iqbar$ and $T_{\qp,h,\q}$ is the splitting matrix of the string+${}^3P_0$ model \cite{Kerbizi:2023luv} describing the splitting $\q\rightarrow h+\qp$. Equation (\ref{eq:wh}) modifies the azimuthal distribution of $h$ produced by Pythia in agreement with the rules of the string+${}^3P_0$ model. It generalizes the recipe in Ref. \cite{Kerbizi:2023cde} by taking into account the spin-entanglement of the $\q\qbar$ pair.

If $h$ is a VM, its decay is performed as in Ref. \cite{Kerbizi:2023cde}, using the spin density matrix $\rho_{aa'}(h)\propto \Trqq\left[\T^{a}_{\qp,h\,\q}\,\rho(\q,\qbar)\T^{a'\,\dagger}_{\qp,h\,\q}\right]$ \cite{Kerbizi:2023luv}, where the splitting amplitude for VM emission is written as $\T_{\qp,h,\q}=\T^a_{\qp,h,\q}\,\V^*_a$ and $\V_a$ is the linear polarization vector of the VM expressed in the QHF. Once $h$ is accepted, the spin correlations are propagated by calculating the joint spin density matrix $\rho(\qp,\qbar)$ of the $\qp\qbar$ pair. It is given by $\rho(\qp,\qbar)\propto \T^{a}_{\qp,h,\q}\,\rho(\q,\qbar)\T^{a'\dagger}\,D_{a'a}$. For a VM emission $D_{a'a}$ is the decay matrix implementing the decay process of the VM \cite{Kerbizi:2021gos,Kerbizi:2023luv}, as required by the Collins-Knowles algorithm \cite{Collins:1987cp,Knowles:1988vs} (Fig. \ref{fig:kinematics}b). For a PS meson emission the indices $a$ and $a'$, and $D_{a',a}$ are removed.

For a splitting from the $\qbar$ side the procedure is analogous, but the splitting amplitude $\T_{\qbarp,H,\qbar}=\Iq\otimes T_{\qbar',H,\qbar}$ is used with $T_{\qbarp,H,\qbar}$ being the antiquark splitting matrix \cite{Kerbizi:2023luv}. If $H$ is emitted, e.g., after the emission of $h$ from the $\q$ side, $H$ is accepted with the probability
\begin{equation}\label{eq:wH}
w_H(\Pt;\ktbar)=\Tr_{\qp\qbar}\left[\T_{\qbarp,H\,\qbar}\,\rho(\qp,\qbar)\T^{\dagger}_{\qbarp,H\,\qbar}\right]/\Tr_{\rm \qp\qbar}\left[\T_{\qbar',H\,\qbar}\,\T^{\dagger}_{\qbarp,H\,\qbar}\right].
\end{equation}
The probability for emitting $H$ is now conditional to the emission of $h$ from $\q$, due to the fact that in Eq. (\ref{eq:wH}) enters the joint spin density matrix $\rho(\qp,\qbar)$. This results in correlations between the azimuthal angles of the transverse momenta $\pt$ and $\Pt$.
For the decay of VMs as well as the propagation of the spin correlations, the same procedure as for the $\q$ splitting above is followed provided that the replacement $\T_{\qp,h,\q}\rightarrow \T_{\qbarp,H,\qbar}$ is performed. The four-momenta of hadrons emitted from the $\qbar$ side are finally expressed in the AHF.

The described procedure is applied until the exit condition of the string fragmentation process is called by Pythia and the process is terminated.

The free parameters of the string+${}^3P_0$ model resposible for the spin effects are the complex mass $\mu=\rm{Re}(\mu)+i\,\rm{Im}(\mu)$ implementing the ${}^3P_0$ mechanism of quark pair production at string breaking, $\fL$ that gives the fraction of longitudinally polarized VMs and $\thetaLT$ allowing for the oblique polarization of VMs \cite{Kerbizi:2021gos}.

\section{Results from simulations of $e^+e^-$ annihilation}\label{sec:results}
Using the developed StringSpinner package, we performed simulations of $e^+e^-\rightarrow \q\qbar\rightarrow \rm{hadrons}$ at the c.m.s energy of $\sqrt{s}=10.6\,\GeV$. It corresponds to the kinematical configuration of the BELLE experiment \cite{Belle:2019nve}. The annihilation reaction is considered to be mediated by a virtual photon, and the allowed flavors are $q=u,d,s$. Thus the production of heavier quarks has been switched off. The parameter settings used for the simulation are the same as those in Ref. \cite{Kerbizi:2023cde}, except for the parameters responsible for spin effects in the production of VMs. In this work, the latter are set to $\fL=0.33$ and $\thetaLT=-\pi/6$ according to a "by eye" tuning found to give satisfactory results also for the SIDIS observables in Ref. \cite{Kerbizi:2023cde}. 

To extract the Collins asymmetries in $e^+e^-$, we consider back-to-back hadrons $h_1$ and $h_2$ produced in the same event. For these hadrons we construct the distribution of the azimuthal angle $\phiH = \phi_1+\phi_2$, with $\phi_i$ being the azimuthal angle of the hadron $i=1,2$ defined with respect to the plane formed by the beam $e^-$ and the axis $\n$. The axis $\n$ can be either the thrust axis $\T$, as in the analysis of Ref. \cite{Belle:2019nve}, or the quark axis $\Q$, as in the analyses of Refs. \cite{Belle:2011cur,BaBar:2013jdt}. The distribution of the produced back-to-back hadrons $h_{1}h_{2}$ is expected to be (see, e.g., Ref. \cite{Boer:2008fr})
\begin{equation}\label{eq:dN}
    N_{12}(\phiH;\Xa\,\Xb)\propto 1+\frac{\sin^2\theta}{1+\cos^2\theta}\,A_{12}(\Xa,\Xb)\,\cos\phiH,
\end{equation}
where the kinematic variable $\Xi$ can be either the fractional energy $z_i=2\,E_i/\sqrt{s}$, with $E_i$ being the energy of $h_i$, or the transverse momentum $\PTi$ with respect to $\n$. Following the analysis in Ref. \cite{Belle:2019nve}, in each kinematic bin the angular distribution in Eq. (\ref{eq:dN}) is used to construct the normalized yield $R_{12}(\phiH;\Xa,\Xb)=N_{12}(\phiH;\Xa,\Xb)/\langle N_{12}(\Xa,\Xb)\rangle$, where $\langle N_{12}(\Xa,\Xb)\rangle$ is the average yield in the considered two-dimensional $\Xa\times \Xb$ bin. Finally normalized yields are constructed for unlike (U) and like (L) charge pairs, and the ratio $R_{12}^{UL}=R_{12}^U/R_{12}^L$ is considered. The latter has a similar expression as in Eq. (\ref{eq:dN}), with the amplitude of the $\cos\phiH$ modulation being given by $A_{12}^{UL}(\Xa,\Xb)\simeq A_{12}^{U}(\Xa,\Xb)-A_{12}^{L}(\Xa,\Xb)$, thus by the difference between the Collins asymmetry for unlike charge hadrons and the asymmetry for like charge hadrons. 

\begin{figure}[tbh]
\centering
\begin{minipage}[b]{0.49\textwidth}
\hspace{-0.8em}
\includegraphics[width=1.0\textwidth]{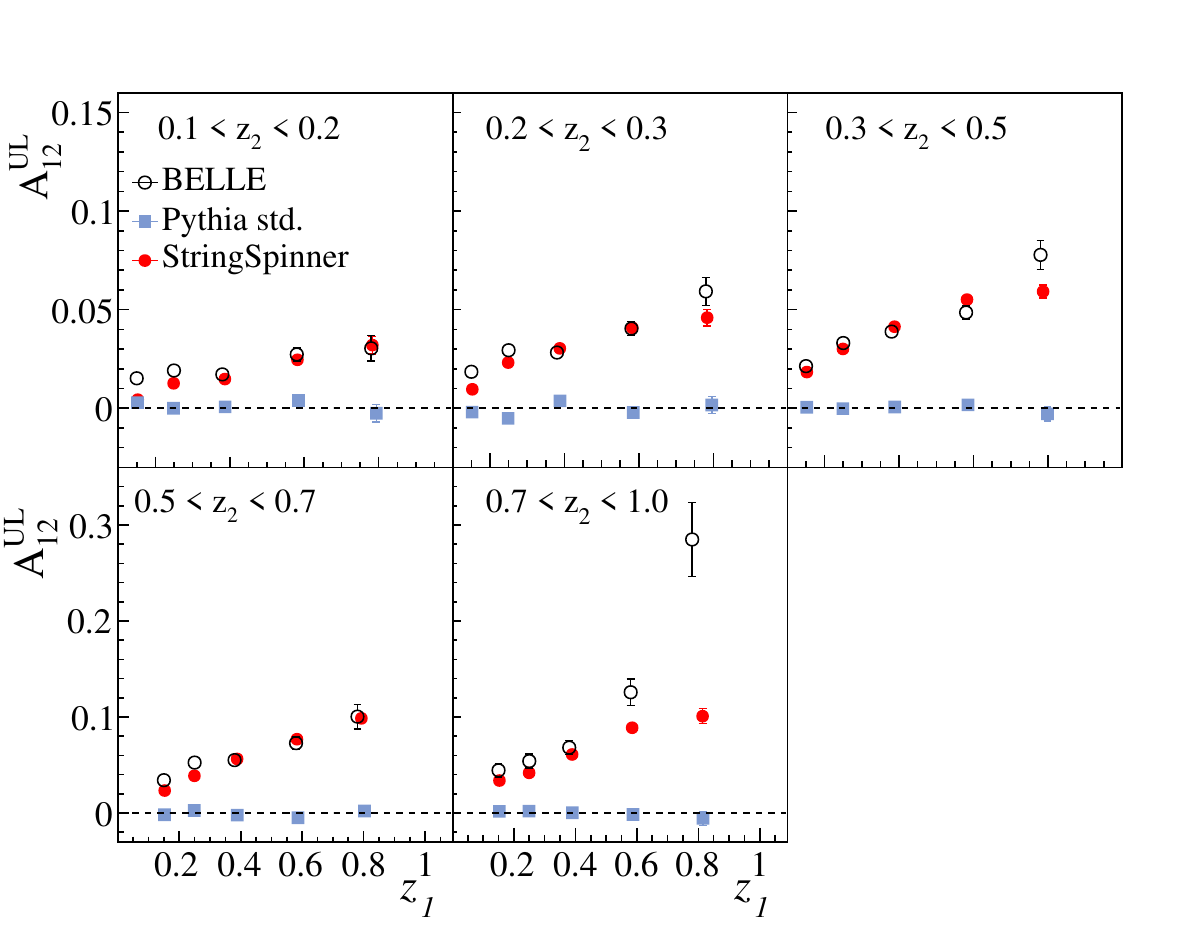}
\vspace{1.8em}
\end{minipage}
\begin{minipage}[b]{0.40\textwidth}
\includegraphics[width=1.0\textwidth]{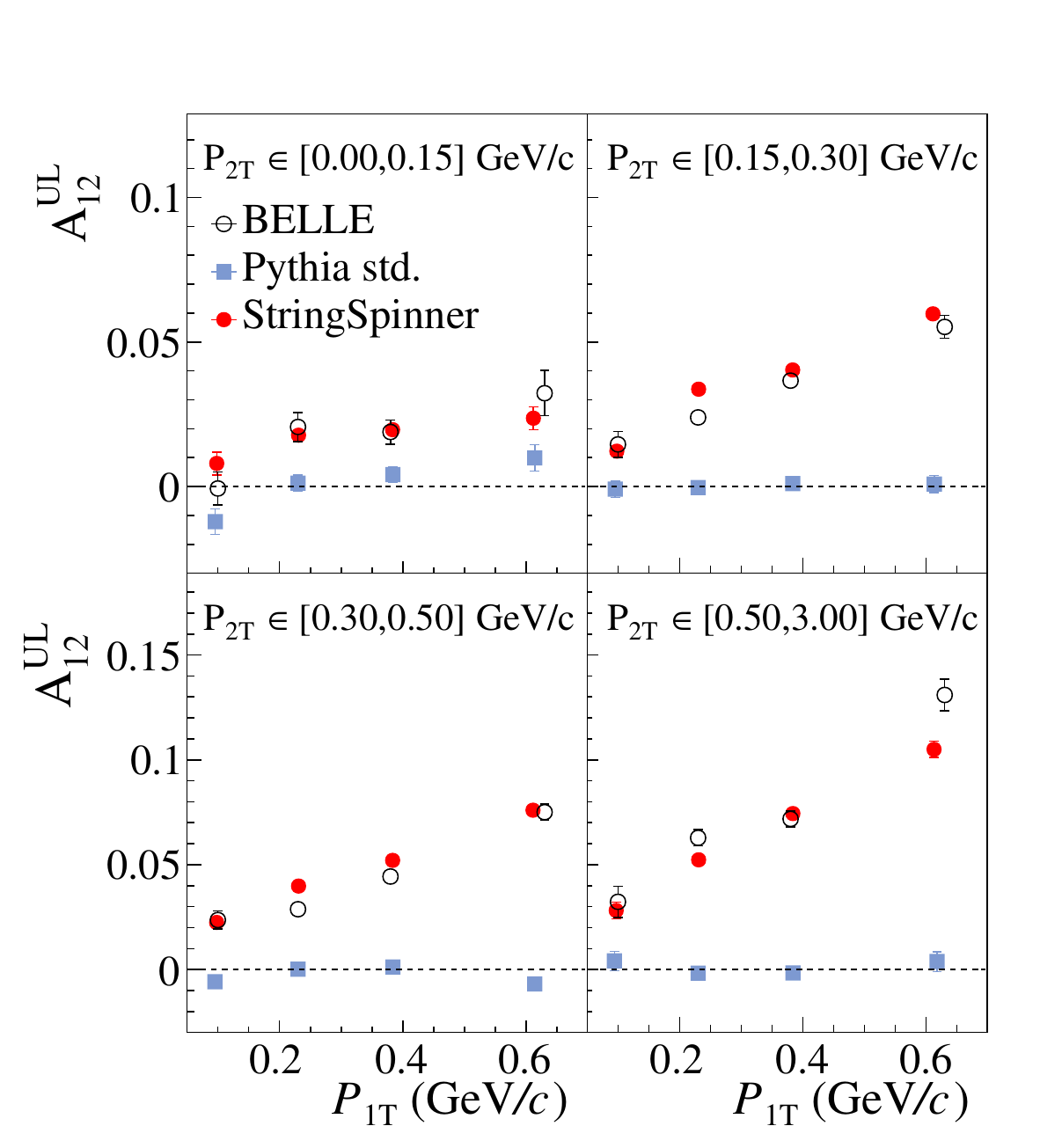}
\end{minipage}
\caption{The Collins asymmetry $\AUL$ for back-to-back $\pi^{\pm}-\pi^{\mp}$ pairs as obtained with StringSpinner (full circles), with standard Pythia (full squares), and as measured by BELLE \cite{Belle:2011cur} (open circles). Left plot: $z_1\times z_2$ binning. Right plot: $\PTa\times \PTb$ binning.}
\label{fig:1}
\end{figure}

Figure \ref{fig:1} shows the comparison between the $\AUL$ asymmetry for back-to-back charged pion pairs obtained from simulations (full points) and measured by BELLE \cite{Belle:2019nve} (open circles). The simulated asymmetry is evaluated using $\n=\T$ and the same kinematic selections as in the BELLE analysis. To obtain the Collins asymmetry in events initiated by $u$, $d$ or $s$ quarks, the BELLE asymmetry is rescaled by $1-f_c(\Xa,\Xb)$, with $f_c(\Xa,\Xb)$ being the fraction of charm-initiated events in the bin $\Xa\times\Xb$ estimated in Ref. \cite{Belle:2019nve}. Thus a vanishing Collins asymmetry for back-to-back charged pions produced in charm-initiated events is assumed, in agreement with Ref. \cite{Belle:2019nve}.

\begin{figure}[tbh]
\centering
\begin{minipage}[b]{0.5\textwidth}
\includegraphics[width=1.0\textwidth]{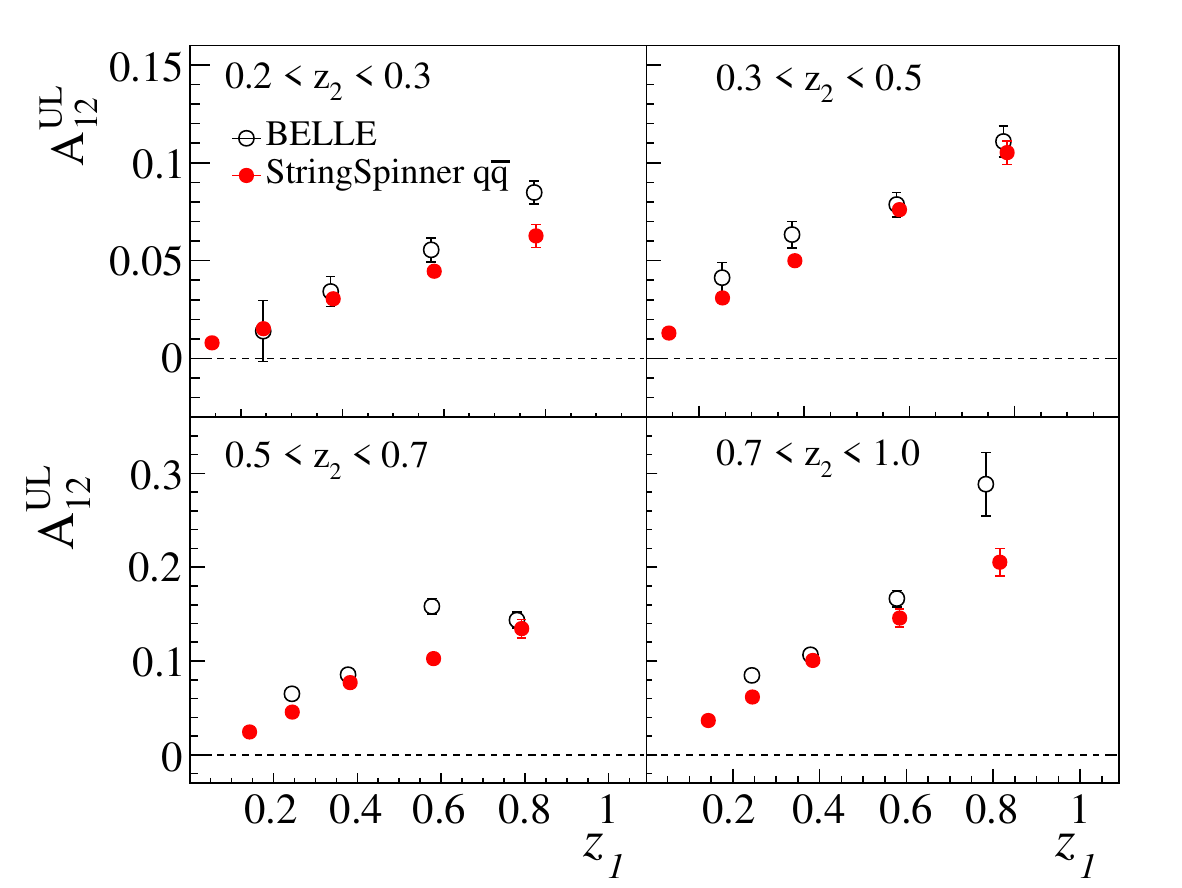}
\end{minipage}
\caption{The Collins asymmetry $\AUL$ for back-to-back $\pi^{\pm}-\pi^{\mp}$ pairs as obtained with StringSpinner (full circles), and as measured by BELLE \cite{Belle:2011cur} (open circles) using as reference axis $\n=\Q$.}
\label{fig:2}
\end{figure}

As can be seen from the left plot in Fig. \ref{fig:2}, which shows the asymmetry as a function of $z_1$ and for selected bins of $z_2$, StringSpinner (closed circles) describes the rising trends with $z$ observed in the measured Collins asymmetry, with the exception of the largest $z_2$ interval where the simulated results are lower. This is not the case with standard Pythia, which gives a vanishing asymmetry (closed squares).

In the right plot in Fig. \ref{fig:1} the simulated Collins asymmetry obtained with StringSpinner is shown as a function of $\PTa$ for the different bins of $\PTb$, and is compared with the corresponding asymmetry as measured by BELLE \cite{Belle:2019nve}. StringSpinner satisfactorily reproduces the rising trends as a function of transverse momentum, as well as the size of the measured asymmetries. Instead, simulations carried with standard Pythia give an asymmetry consistent with zero. From a deeper investigation of the StringSpinner result, it turns out that the linear $\PT$-dependence of the $\AUL$ asymmetry is an effect of the misalignment between the thrust axis $\T$ and the quark axis $\Q$. It is in fact no longer linear if $\n=\Q$.

Using StringSpinner with the same parameter settings we evaluated also the Collins asymmetries for back-to-back $\pi^0-\pi^{\pm}$ pairs and $\eta-\pi^{\pm}$ pairs. A similar description of the data \cite{Belle:2019nve} as in Fig. \ref{fig:1} was found.

Finally, in Fig. \ref{fig:2} is shown the $\AUL$ asymmetry as a function of $z_1$ and in bins of $z_2$ obtained with StringSpinner (full circles) using $\n=\Q$. The $z_1\times z_2$ binning as well as the kinematic selections are the same as in the BELLE analysis in Ref. \cite{Belle:2011cur}. The corresponding asymmetry as measured by BELLE \cite{Belle:2011cur} (empty points) is also shown. As can be seen, StringSpinner provides a satisfactory description of the data also in the case when the $\AUL$ asymmetry is evaluated taking as reference axis the quark axis.

\section{Conclusions}\label{sec:conclusions}
The string+${}^3P_0$ model of hadronization is implemented for the first time in the Pythia 8 event generator for the simulation of the $e^+e^-$ annihilation to hadrons with quark spin effects. For this purpose, the recently proposed recursive recipe in Ref. \cite{Kerbizi:2023luv} for the polarized string fragmentation of a quark pair with entangled spin states has been used. The implementation in Pythia is performed by a new development of the StringSpinner package. The new package is used to simulate $e^+e^-$ annihilation events at the c.m.s energy $\sqrt{s}=10.6\,\GeV$, assuming that the annihilation occurs by the exchange of a virtual photon. The Collins asymmetries for back-to-back hadrons have been calculated and compared to the data by the BELLE experiment, finding a satisfactory agreement.

More phenomenological studies are foreseen, e.g. the comparison with the Collins asymmetries measured by the BABAR and BESS III experiments, and the new StringSpinner package is expected to be made public soon.

\subsection*{Acknowledgments}
The work of AK was supported by the Ministry of University and Research (MUR) within the POLFRAG project, CUP n. J97G22000510001.


\begin{thebibliography}{99}

\bibitem{Collins:1981uk}
J.~C.~Collins and D.~E.~Soper,
\href{https://doi.org/10.1016/0550-3213(81)90339-4}{Nucl. Phys. B \textbf{193} (1981), 381}
[erratum: Nucl. Phys. B \textbf{213} (1983), 545].

\bibitem{Anselmino:2020vlp}
M.~Anselmino, A.~Mukherjee and A.~Vossen,
\href{https://doi.org/10.1016/j.ppnp.2020.103806}{Prog. Part. Nucl. Phys. \textbf{114} (2020), 103806.}

\bibitem{Belle:2019nve}
H.~Li \textit{et al.} [Belle],
\href{https://doi.org/10.1103/PhysRevD.100.092008}{Phys. Rev. D \textbf{100} (2019) no.9, 092008}.

\bibitem{Belle:2011cur}
A.~Vossen \textit{et al.} [Belle],
\href{https://doi.org/10.1103/PhysRevLett.107.072004}{Phys. Rev. Lett. \textbf{107} (2011), 072004}.

\bibitem{BaBar:2013jdt}
J.~P.~Lees \textit{et al.} [BaBar],
\href{https://doi.org/10.1103/PhysRevD.90.052003}{Phys. Rev. D \textbf{90} (2014) no.5, 052003.}

\bibitem{BaBar:2015mcn}
J.~P.~Lees \textit{et al.} [BaBar],
\href{https:/doi.org/10.1103/PhysRevD.92.111101}{Phys. Rev. D \textbf{92} (2015) no.11, 111101}.

\bibitem{BESIII:2015fyw}
M.~Ablikim \textit{et al.} [BESIII],
\href{https://doi.org/10.1103/PhysRevLett.116.042001}{Phys. Rev. Lett. \textbf{116} (2016) no.4, 042001.}

\bibitem{Bierlich:2022pfr}
C.~Bierlich, S.~Chakraborty, N.~Desai, \textit{et al.},
\href{https://doi.org/10.21468/SciPostPhysCodeb.8}{SciPost Phys. Codebases 8 (2022).}

\bibitem{Kerbizi:2021gos}
A.~Kerbizi, X.~Artru and A.~Martin,
\href{https://doi.org/10.1103/PhysRevD.104.114038}{Phys. Rev. D \textbf{104} (2021) no.11, 114038.}


\bibitem{Kerbizi:2023luv}
A.~Kerbizi and X.~Artru,
\href{https://arxiv.org/abs/2312.14694}{arXiv:2312.14694 [hep-ph]}.

\bibitem{Kerbizi:2021pzn}
A.~Kerbizi and L.~L\"onnblad,
\href{https://doi.org/10.1016/j.cpc.2021.108234}{Comput. Phys. Commun. \textbf{272} (2022), 108234}.

\bibitem{Kerbizi:2023cde}
A.~Kerbizi and L.~L\"onnblad,
\href{https://doi.org/10.1016/j.cpc.2023.108886}{Comput. Phys. Commun. \textbf{292} (2023), 108886}.

\bibitem{Collins:1987cp}
J.~C.~Collins,
\href{https://doi.org/10.1016/0550-3213(88)90654-2}{Nucl. Phys. B \textbf{304} (1988), 794-804.}

\bibitem{Knowles:1988vs}
I.~G.~Knowles,
\href{https://doi.org/10.1016/0550-3213(88)90092-2}{Nucl. Phys. B \textbf{310} (1988), 571-588.}

\bibitem{Boer:2008fr}
D.~Boer,
\href{https://doi.org/10.1016/j.nuclphysb.2008.06.011}{Nucl. Phys. B \textbf{806} (2009), 23-67}.

\end{thebibliography}

\end{document}